\documentclass[runinaddress,showpacs,twocolumn,prb]{revtex4}
\usepackage{amsmath}
\usepackage{amssymb}
\usepackage{graphicx}

\begin{document}

\title{Spin transverse force and intrinsic quantum transverse transport}
\author{Bin Zhou$^{1,2}$, Li Ren$^{3}$, and Shun-Qing Shen$^{1}$}
\affiliation{$^{1}$Department of Physics, and Center for Theoretical and Computational
Physics, The University of Hong Kong, Pokfulam Road, Hong Kong, China\\
$^{2}$Department of Physics, Hubei University, Wuhan 430062, China\\
$^{3}$Department of Physics and Research Center of Quantum Manipulation,
Fudan University, 220 Handan Road, Shanghai 200433, China}
\date{\today }

\begin{abstract}
The spin-orbit coupling may generate spin transverse force on moving
electron spin, which gives a heuristic picture for the quantum transverse
transport of electron. A relation between the spin and anomalous Hall
conductance and spin force was established, and applied to several systems.
It was predicted that the sign change of anomalous Hall conductance can
occur in diluted magnetic semiconductors of narrow band and can be applied
to identify intrinsic mechanism experimentally.
\end{abstract}

\pacs{72.25.Dc, 72.20.-i, 72.10.-d, 85.75.-d}
\maketitle

\section{Introduction}

The theory of anomalous Hall effect has a long history since 1950s.\cite{AHE}
It was realized that several different mechanisms are contributed to the
total Hall conductance in ferromagnetic metals and semiconductors. Except
for the skew scattering and the side jump from the impurity scattering,\cite%
{Karplus54pr,Luttinger58pr,Smit58p,Berger70prb} an intrinsic mechanism tells
that the spin-orbit coupling in the electronic band structure of the system
may induce a non-zero Berry phase or magnetic monopole in the momentum space
and cannot be neglected in the transverse transport of electrons, especially
in the diluted magnetic semiconductors.\cite{Jungwirth02prl,Onada02jpsj} On
the other hand, the spin aspect of transverse transport was also studied
recently, in which an external electric field may drive electrons to form a
transverse spin current in the systems with spin-orbit coupling even in
paramagnetic electronic systems.\cite{Murakami03sci,Sinova04prl,Shen04prl}
These two effects reflect the charge and spin aspects of electron transport,
respectively, and have some common features as their physical origin stems
from the same spin-orbit coupling of conduction electrons. More and more
experiments support the intrinsic mechanism.\cite%
{Fang04Science,Lee04sci,Wunderlich05prl} Since the spin-orbit coupling
reflects the interaction between the electron spin, momentum and external or
environmental potential, it was shown that this coupling, as an extension of
Ehrenfest's theorem in quantum mechanics, may generate a spin transverse
force on the spin current instead of the Lorentz force on electric current
in a magnetic field in conventional Hall effect.\cite{Shen05prl} This spin
transverse force provides a new route to get insight of the transverse
motion of electrons.

In this paper we focus on the relation between the spin force and intrinsic
quantum transverse transport. For the electronic system with spin-orbit
coupling and exchange coupling, it was found that a spin transverse force
exerts on moving electron spins in the anomalous transverse effect, as a
result, a set of formula based on spin force were derived for intrinsic spin
and anomalous Hall conductance. As an intrinsic feature it was predicted
that the sign of the anomalous Hall conductance can be changed when the
system breaks both the bulk and structural inversion symmetry and time
reversal symmetry. It is also shown that spin Hall effect in the
two-dimensional (2D) Luttinger system is equivalent to the intrinsic
anomalous and spin Hall problems in a ferromagnetic metal, and its
robustness against impurities can be also understood very well from the
point of view of spin force.

\section{General formula}

To develop a general formula for spin and anomalous Hall conductance, we
start with an effective Hamiltonian for electrons with spin 1/2,
\begin{equation}
H=\epsilon (p)+\sum_{\alpha =x,y,z}d_{\alpha }(p)\sigma _{\alpha },
\label{general}
\end{equation}%
where $\epsilon (p)=p^{2}/2m^{\ast }$ is the kinetic energy with the band
electron effective mass $m^{\ast }$ and $\sigma _{\alpha }$ are the Pauli
matrices. $d_{\alpha }(p)$ are the the momentum-dependent coefficients which
describes the spin-orbit interactions and exchange interaction of magnetic
impurities. The energy eigenvalues are $E_{pn}=$ $\epsilon (p)+\mu d(p)$
with $\mu =\pm ,$ and $d=\sqrt{d_{x}^{2}+d_{y}^{2}+d_{z}^{2}}$, and the
corresponding states are denoted by $\left\vert p,\mu \right\rangle $. Using
the Heisenberg equation, the kinetic velocity operator ($j=x,y,z$) is
defined as
\begin{equation}
v_{j}=\frac{1}{i\hbar }[r_{j},H]=\frac{p_{j}}{m^{\ast }}+\sum_{\alpha =x,y,z}%
\frac{\partial d_{\alpha }}{\partial p_{j}}\sigma _{\alpha },
\label{velocity}
\end{equation}%
where the first term is the canonical velocity, and second part can be
regarded as the spin gauge field or anomalous velocity, $\left( e/m^{\ast
}c\right) \mathcal{A}_{j}=\sum_{\alpha =x,y,z}\partial d_{\alpha }/\partial
p_{j}\sigma _{\alpha }$. As an extension of the Ehrenfest's theorem,\cite%
{Shen05prl} the spin-dependent force is introduced as the derivative of
kinetic momentum operator with the respect to time,%
\begin{equation}
F_{j}=\frac{m^{\ast }}{i\hbar }[v_{j},H]=\frac{2m^{\ast }}{\hbar }\epsilon
_{\alpha \beta \gamma }\frac{\partial d_{\alpha }}{\partial p_{j}}d_{\beta
}\sigma _{\gamma }.
\end{equation}%
This spin force is a purely quantum quantity, and has no classical
counterpart.

The quantum transverse transport of electrons caused by a weak electric
field $\mathbf{E}$ can be calculated within the framework of linear response
theory. The Kubo formula for the d.c. conductivity gives,\cite{Mahan}
\begin{equation}
\sigma _{ij}=\frac{e^{2}\hbar }{\Omega }\sum_{p,\mu \neq \mu ^{\prime }}%
\frac{\left( f_{p\mu }-f_{p\mu ^{\prime }}\right) \text{Im}\left(
\left\langle p\mu \right\vert v_{i}\left\vert p\mu ^{\prime }\right\rangle
\left\langle p\mu ^{\prime }\right\vert v_{j}\left\vert p\mu \right\rangle
\right) }{\left( E_{p\mu }-E_{p\mu ^{\prime }}\right) \left( E_{p\mu
}-E_{p\mu ^{\prime }}+i\delta ^{+}\right) },
\end{equation}%
with $\delta ^{+}\rightarrow 0^{+}$, and the Dirac-Fermi distribution
function $f_{p,\mu }=1/\left\{ \exp [\beta \left( E_{\mu }(p)-\mu \right)
]+1\right\} .$ With the formula of the kinetic velocity in Eq.(\ref{velocity}%
)
\begin{equation}
\left\langle p\mu \right\vert v_{i}\left\vert p\mu ^{\prime }\right\rangle
=\sum_{\alpha =x,y,z}\frac{\partial d_{\alpha }}{\partial p_{j}}\left\langle
p\mu \right\vert \sigma _{\alpha }\left\vert p\mu ^{\prime }\right\rangle
\end{equation}%
for $\mu \neq \mu ^{\prime }$ and furthermore an identity can be proved for
this system,
\begin{equation}
\text{Im}\left( \left\langle p\mu \right\vert \sigma _{\alpha }\left\vert
p\mu ^{\prime }\right\rangle \left\langle p\mu ^{\prime }\right\vert \sigma
_{\beta }\left\vert p\mu \right\rangle \right) =\mu \epsilon _{\alpha \beta
\gamma }\frac{d_{\gamma }}{d},
\end{equation}%
for $\mu \neq \mu ^{\prime }$. We limit our discussion to the case that two
spectra are non-degenerated in the whole momentum space, $i.e.$ , $d\neq 0$
in the whole momentum $p$ space such that $\delta ^{+}$ can be taken to be
zero before the integral over $p$.\cite{singularity} Thus the intrinsic
transverse conductance can be expressed as
\begin{equation}
\sigma _{ij}=-\frac{e^{2}\hbar }{2\Omega }\sum_{p}\frac{\left(
f_{p,-}-f_{p,+}\right) }{d^{3}}\epsilon _{\alpha \beta \gamma }\frac{%
\partial d_{\alpha }}{\partial p_{i}}\frac{\partial d_{\beta }}{\partial
p_{j}}d_{\gamma }.
\end{equation}%
This recovers the conductance formula in terms of the Berry curvature.\cite%
{Qi05xxx,Bernevig05prb,Sundaram99prb} On the other hand, the anticommutators
of the spin force $F_{j}$ and spin gauge field $\mathcal{A}_{i}$ gives
\begin{equation}
\text{Tr}\left[ \left\{ F_{j},\mathcal{A}_{i}\right\} \right] =\frac{%
8m^{\ast 2}c}{\hslash e}\epsilon _{\alpha \beta \gamma }\frac{\partial
d_{\alpha }}{\partial p_{i}}\frac{\partial d_{\beta }}{\partial p_{j}}%
d_{\gamma },
\end{equation}%
where the trace runs through the spin variables, $\sum_{\mu }\left\langle
p\mu \right\vert \cdots \left\vert p\mu \right\rangle $. In this way we
established a relation between the electric conductance and spin force,
\begin{equation}
\sigma _{ij}=-\frac{e^{3}\hbar ^{2}}{16m^{\ast 2}c\Omega }\sum_{p}\frac{%
\left( f_{p,-}-f_{p,+}\right) }{d^{3}}\text{Tr}\left[ \{F_{j},\mathcal{A}%
_{i}\}\right] .  \label{ahc}
\end{equation}%
Since Tr$\left[ \{F_{j},v_{i}\}\right] =\frac{e}{m^{\ast }c}$Tr$\left[
\{F_{j},\mathcal{A}_{i}\}\right] ,$ the anomalous Hall conductance is
determined by both the spin force and the spin gauge field, or anomalous
part of the velocity. This fact reflects the physical origin of spin-orbit
coupling in this effect. Note that Tr$\left[ \{F_{j},\mathcal{A}_{i}\}\right]
=-$ Tr$\left[ \{F_{i},\mathcal{A}_{j}\}\right] $, so $\sigma _{ij}=-\sigma
_{ji}$ for $i\neq j$, that is the so-called Onsager relation. It is noticed
that Tr$\left[ \left\{ F_{j},\mathcal{A}_{i}\right\} \right] =0$ if any one
of $d_{\alpha }$ is zero.

The Kubo formula for spin Hall conductance is written as%
\begin{equation}
\sigma _{ij}^{\gamma }=\frac{e\hslash }{\Omega }\sum_{p,\mu \neq \mu
^{\prime }}\frac{\left( f_{p\mu }-f_{p\mu ^{\prime }}\right) \text{Im}%
\left\langle p\mu \right\vert J_{i}^{\gamma }\left\vert p\mu ^{\prime
}\right\rangle \left\langle p\mu ^{\prime }\right\vert v_{j}\left\vert p\mu
\right\rangle }{\left( E_{p\mu }-E_{p\mu ^{\prime }}\right) \left( E_{p\mu
}-E_{p\mu ^{\prime }}+i\delta^{+} \right) },  \label{Kubospin}
\end{equation}%
where spin current operator $J_{i}^{\gamma }$ is defined conventionally as $%
J_{i}^{\gamma }=\left( \hbar /4\right) \{v_{i},\sigma _{\gamma }\}$, note
that this choice a natural one but not a unique one in the presence of
spin-orbit coupling since these is no continuity equation for spin density
as is the case of for charge density. Following the calculation mentioned
above, from Eq.(\ref{Kubospin}) one can obtain \cite{Bernevig05prb}
\begin{equation}
\sigma _{ij}^{\gamma }=\frac{e\hslash ^{2}}{4\Omega }\sum \frac{\left(
f_{p,-}-f_{p,+}\right) }{d^{3}}\frac{\partial \epsilon }{\partial p_{i}}%
d_{\alpha }\frac{\partial d_{\beta }}{\partial p_{j}}\epsilon _{\alpha \beta
\gamma }.
\end{equation}%
Similarly, the spin conductance can be also expressed in terms of the spin
force,%
\begin{equation}
\sigma _{ij}^{\gamma }=-\frac{e\hslash ^{2}}{16m^{\ast }\Omega }\sum_{p}%
\frac{\left( f_{p,-}-f_{p,+}\right) }{d^{3}}\text{Tr}\left[
\{F_{j},J_{i}^{\gamma }\}\right] ,  \label{spinhall}
\end{equation}%
which is given by the anticommutator of the spin force and spin current
operators $J_{i}^{\gamma }$ with%
\begin{equation}
\text{Tr}\left[ \{F_{j},J_{i}^{\gamma }\}\right] =-4m^{\ast }\frac{\partial
\epsilon }{\partial p_{i}}d_{\alpha }\frac{\partial d_{\beta }}{\partial
p_{j}}\epsilon _{\alpha \beta \gamma }.
\end{equation}

On the other hand, due to the spin-orbit coupling, an electric field can
also induce a non-zero spin polarization. Its linear response to the field
is,
\begin{eqnarray}
\chi _{j}^{\gamma } &=&\frac{\hbar }{2}\left( \left\langle \sigma _{\gamma
}\right\rangle _{E}-\left\langle \sigma _{\gamma }\right\rangle
_{E=0}\right) /E_{j}  \notag \\
&=&-\frac{e\hbar ^{3}}{32m^{\ast }\Omega }\sum_{p}\frac{\left(
f_{p,-}-f_{p,+}\right) }{d^{3}}\text{Tr}\left[ \{F_{j},\sigma _{\gamma }\}%
\right] .
\end{eqnarray}%
In this way we have established an explicit relation between the spin force
and the quantum spin and charge transverse transport in a system with
spin-orbit coupling.

In general, for any observable $O$, its linear response to an external
electric field $\mathbf{E}$, $\left\langle O_{i}\right\rangle =\chi
_{ij}E_{j}$, where, from the Kubo formula,
\begin{equation}
\chi _{ij}=-\frac{e\hslash ^{2}}{16m^{\ast }\Omega }\sum_{p}\frac{\left(
f_{p,-}-f_{p,+}\right) }{d^{3}}\text{Tr}\left[ \{F_{j},O_{i}\}\right] .
\end{equation}

\section{Applications of the general formula}

\subsection{2D ferromagnetic system with the \textit{Wurtzite }and\textit{\
zinc-blende }structures}

Now we apply the general formula to several systems in semiconductors. We
first consider an effective Hamiltonian for a two-dimensional ferromagnetic
system with the \textit{Wurtzite }and\textit{\ zinc-blende }structures,%
\begin{equation}
H=\frac{p^{2}}{2m^{\ast }}+a_{0}f(p)(p_{y}\sigma _{x}-p_{x}\sigma
_{y})+h_{0}\sigma _{z}.  \label{f}
\end{equation}%
The exchange field due to the magnetic impurities or Coulomb interaction is
taken to be uniform, as an mean field effect, and is normal to the plain. $%
f(p)=1$ for the Wurtzite structure, and $f(p)=p^{2}$ for the zinc-blende
structure such as Hg$_{1-x}$Mn$_{x}$Te.\cite{Culcer03prb} The components of
spin gauge field $\mathcal{A}$ in this system are%
\begin{equation}
\mathcal{A}_{x}=\frac{m^{\ast }ca_{0}}{e}\left[ p_{y}\frac{\partial f}{%
\partial p_{x}}\sigma _{x}-\left( f+p_{x}\frac{\partial f}{\partial p_{x}}%
\right) \sigma _{y}\right] ,
\end{equation}%
\begin{equation}
\mathcal{A}_{y}=\frac{m^{\ast }ca_{0}}{e}\left[ \left( f+p_{y}\frac{\partial
f}{\partial p_{y}}\right) \sigma _{x}-p_{x}\frac{\partial f}{\partial p_{y}}%
\sigma _{y}\right] ,
\end{equation}%
and the spin force is,%
\begin{equation}
\mathbf{F}_{s}=\frac{4m^{\ast }}{\hbar ^{2}}a_{0}^{2}f^{2}J_{s}^{z}\times
\hat{z}-\frac{2m^{\ast }a_{0}h_{0}}{\hbar }\nabla _{p}(f\mathbf{p}\cdot
\mathbf{\sigma }).
\end{equation}%
From the formula we notice that a spin transverse force arises due to the
spin-orbit coupling if there exists a spin current along the electric field
and also an in-plain spin force arises due to the interference between the
spin-orbit coupling and the exchange field. To calculate the intrinsic Hall
conductance, we notice that the anticommutator,
\begin{equation}
\text{Tr}\left[ \{F_{y},\mathcal{A}_{x}\}\right] =\frac{8m^{\ast 2}ch_{0}}{%
\hbar e}a_{0}^{2}f\left[ d\left( pf\right) /dp\right] ,
\end{equation}%
which vanishes when $h_{0}=0$. Thus if the system does not violate the time
reversal symmetry no anomalous Hall current circulates. Assuming at zero
temperature and taking the bottom subband to be occupied and the top one to
be empty. The anomalous Hall conductance is
\begin{equation}
\sigma _{xy}=-\frac{e^{2}}{2h}\left( 1-h_{0}/\left[
a_{0}^{2}x_{f}^{2}+h_{0}^{2}\right] ^{1/2}\right) ,
\end{equation}%
where $x_{f}^{2}=p_{F}^{2}f^{2}$ at the Fermi surface. If $%
a_{0}p_{F}f>>h_{0} $, the anomalous Hall conductance is almost quantized, $%
\sigma _{xy}\simeq -\frac{e^{2}}{2h}.$This is consistent with the theory of
Berry curvature.\cite{Culcer03prb}

According to the Eq.(\ref{spinhall}), one can obtain the intrinsic spin Hall
conductance
\begin{equation}
\sigma _{xy}^{z}=\frac{e\hslash ^{2}}{4m^{\ast }\Omega }\sum_{p}\frac{\left(
f_{p,-}-f_{p,+}\right) p_{x}^{2}a_{0}^{2}f^{2}}{\left[
p^{2}a_{0}^{2}f^{2}+h_{0}^{2}\right] ^{3/2}}.
\end{equation}%
For the Wurtzite structure, i.e., $f(p)=1$, assuming the Fermi energy $\mu $
lies in the gap, the intrinsic spin Hall conductance is
\begin{equation}
\sigma _{xy}^{z}=\frac{e\hslash ^{2}}{16m^{\ast }\pi }\frac{\left(
h_{0}-\eta _{1}\right) ^{2}}{a_{0}^{2}\eta _{1}},
\end{equation}%
where%
\begin{equation}
\eta _{1}=\left( h_{0}^{2}+p_{F}^{2}a_{0}^{2}\right) ^{1/2},
\end{equation}%
with%
\begin{eqnarray}
p_{F} &=&\sqrt{2}\{\left[ m^{\ast 2}\left( a_{0}^{4}m^{\ast
2}+h_{0}^{2}+2a_{0}^{2}\mu m^{\ast }\right) \right] ^{1/2}  \notag \\
&&+m^{\ast }\left( a_{0}^{2}m^{\ast }+\mu \right) \}^{1/2}.
\end{eqnarray}%
For the zinc-blende structure, i.e., $f(p)=p^{2}$,
\begin{equation}
\sigma _{xy}^{z}=\frac{e\hslash ^{2}}{32m^{\ast }\pi }\frac{\left(
h_{0}-\eta _{2}\right) ^{2}}{a_{0}^{2}\eta _{2}},
\end{equation}%
where
\begin{equation}
\eta _{2}=\left( h_{0}^{2}+p_{F}^{4}a_{0}^{2}\right) ^{1/2},
\end{equation}%
with
\begin{eqnarray}
p_{F} &=&\{2m^{\ast }(\mu +\left[ h_{0}^{2}-4a_{0}^{2}m^{\ast 2}\left(
h_{0}^{2}-\mu ^{2}\right) \right] ^{1/2})  \notag \\
&&/\left( 1-4a_{0}^{2}m^{\ast 2}\right) \}^{1/2}.
\end{eqnarray}

\subsection{2D ferromagnetic electrons gas with both Rashba and Dresselhaus
coupling}

In some materials the spin orbit coupling has two different contributions
when a system has the structural and bulk inversion asymmetry. Here we
consider a two-dimensional ferromagnetic electrons gas with both Rashba and
Dresselhaus coupling,
\begin{equation}
H=\frac{p^{2}}{2m^{\ast }}-\lambda \left( p_{x}\sigma _{y}-p_{y}\sigma
_{x}\right) -\beta \left( p_{x}\sigma _{x}-p_{y}\sigma _{y}\right)
+h_{0}\sigma _{z}.  \label{RD}
\end{equation}%
In the language of $d_{\alpha }$ in Eq.(\ref{general}),
\begin{equation}
d_{x}=\lambda p_{y}-\beta p_{x,}
\end{equation}%
\begin{equation}
d_{y}=-\lambda p_{x}+\beta p_{y},
\end{equation}%
\begin{equation}
d_{z}=h_{0}.
\end{equation}%
It is straightforward that
\begin{equation}
\text{Tr}\left[ \{F_{y},\mathcal{A}_{x}\}\right] =\frac{8m^{\ast 2}ch_{0}}{%
\hbar e}\left( \lambda ^{2}-\beta ^{2}\right) ,
\end{equation}%
which is independent of $p$. Using the formula for the Hall conductance, we
conclude immediately that the anomalous Hall conductance vanishes at $%
h_{0}=0 $. The Hall conductance will change its sign near the point of $%
\lambda ^{2}=\beta ^{2}$. Since $h_{0}\neq 0,$ the band spectra opens a
finite gap at $p=0,$ and degeneracy of two electron bands is removed. If one
assume that the Fermi surface is below the the up-band of the spectrum, the
anomalous Hall conductance $\sigma _{xy}$ reads
\begin{equation}
\sigma _{xy}=-\frac{e^{2}}{2h}\text{sgn}\left( \lambda ^{2}-\beta
^{2}\right) \left( 1-\gamma \right) ,
\end{equation}%
where%
\begin{equation}
\gamma =\frac{1}{2\pi }\int_{0}^{2\pi }d\phi \frac{\left\vert \lambda
^{2}-\beta ^{2}\right\vert }{\Pi \left( \phi \right) \left[ 1+\text{ }%
p_{F}^{2}(\phi )\Pi (\phi )/h_{0}^{2}\right] ^{1/2}},
\end{equation}%
with
\begin{equation}
\Pi \left( \phi \right) =\lambda ^{2}+\beta ^{2}-2\lambda \beta \sin 2\phi ,
\end{equation}%
and $p_{F}$ is the Fermi momentum,
\begin{eqnarray}
p_{F} &=&\sqrt{2m^{\ast }}\{\mu +m^{\ast }\lambda \beta \left( \lambda \beta
-2\sin 2\phi \right) +[h_{0}^{2}-\mu ^{2}  \notag \\
&&+\left( m^{\ast }\lambda ^{2}\beta ^{2}+\mu -2m^{\ast }\lambda \beta \sin
2\phi \right) ^{2}]^{1/2}\}^{1/2}.
\end{eqnarray}%
The dependence $\sigma _{xy}$ on $\beta $ of numerical calculations is
plotted in Fig.1. The parameters are taken as \cite{Culcer03prb}: $\lambda
=23$ meV nm/$\hbar $, $h_{0}=1.38$ meV, $\mu =1.38$ meV, $m^{\ast }=0.9m_{0}$%
, and $\beta $ is comparable with $\lambda .$ It is shown that the anomalous
Hall conductance will change its sign near the point of $\lambda ^{2}=\beta
^{2}$. If $p_{F}\left\vert \lambda -\beta \right\vert \gg h_{0}$, $\sigma
_{xy}$ is approximately equal to $-$sgn$\left( \lambda ^{2}-\beta
^{2}\right) \frac{e^{2}}{2h}$.

If one assume that the Fermi surface is below the the up-band of the
spectrum, the intrinsic spin Hall conductance can be also obtained
\begin{equation}
\sigma _{xy}^{z}=\frac{\hslash ^{2}e}{4m}\frac{1}{\left( 2\pi \right) ^{2}}%
\int_{0}^{2\pi }d\phi \frac{\left( \lambda ^{2}-\beta ^{2}\right) \left(
h_{0}-\eta \left( \phi \right) \right) ^{2}\cos ^{2}\phi }{\Pi \left( \phi
\right) ^{2}\eta \left( \phi \right) }.
\end{equation}%
While the Fermi energy is above the gap, the intrinsic spin Hall conductance
is%
\begin{eqnarray}
\sigma _{xy}^{z} &=&\frac{\hslash ^{2}e}{4m}\frac{1}{\left( 2\pi \right) ^{2}%
}\int_{0}^{2\pi }d\phi \frac{\left( \lambda ^{2}-\beta ^{2}\right) \cos
^{2}\phi }{\Pi \left( \phi \right) ^{2}\eta _{-}\eta _{+}}  \notag \\
&&\times \left( h_{0}^{2}-\eta _{-}\eta _{+}\right) \left( \eta _{+}-\eta
_{-}\right) ,
\end{eqnarray}%
with
\begin{equation}
\eta _{\pm }=\left[ h_{0}^{2}+\text{ }p_{\pm }^{2}\Pi \left( \phi \right) %
\right] ^{1/2},
\end{equation}%
and%
\begin{eqnarray}
p_{\pm } &=&\sqrt{2m^{\ast }}\{\mu +m^{\ast }\lambda \beta \left( \lambda
\beta -2\sin 2\phi \right) \mp \lbrack h_{0}^{2}-\mu ^{2}  \notag \\
&&+\left( m^{\ast }\lambda ^{2}\beta ^{2}+\mu -2m^{\ast }\lambda \beta \sin
2\phi \right) ^{2}]^{1/2}\}^{1/2}.
\end{eqnarray}%
The numerical calculation of the intrinsic spin Hall conductance $\sigma
_{xy}^{z}$ with $\beta $ is plotted by the dashed line in Fig.1. When sgn$%
\left( \lambda ^{2}-\beta ^{2}\right) =+1$, the intrinsic spin Hall
conductance $\sigma _{xy}^{z}$ is positive; sgn$\left( \lambda ^{2}-\beta
^{2}\right) =-1$, $\sigma _{xy}^{z}$ is negative.

\begin{figure}[tbp]
\includegraphics[width=8cm]{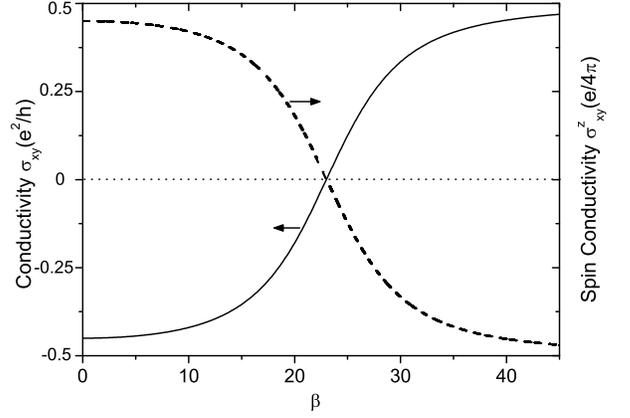}
\caption{Variations of the conductivity $\protect\sigma _{xy}$ and the spin
conductivity $\protect\sigma _{xy}^{z}$ with the coupling parameter $\protect%
\beta $ (meV nm/$\hbar $). The parameters are $\protect\lambda =23$ meV nm/$%
\hbar $, $h_{0}=1.38$ meV, $\protect\mu =1.38$ meV, and $m^{\ast }=0.9m_{0}$%
. The solid line corresponds to $\protect\sigma _{xy}$, and the dashed line
to $\protect\sigma _{xy}^{z}$.}
\end{figure}

The sign change of anomalous Hall conductance as that of spin Hall
conductance \cite{Shen04prb} may be observed experimentally in some diluted
magnetic GaAs quantum wells where the Rashba and Dresselhaus coupling are
usually of the same order of magnitude. The Rashba coupling is adjustable by
a gate field perpendicular to the electron gas plane. Thus there is no
practical difficulty to achieve the situation near $\lambda =\pm \beta $.
Since the Rashba coupling has a relation to the gate voltage, $\lambda
=\lambda _{0}+\delta \lambda V_{g}$, the gate field can be used to control
the direction of intrinsic anomalous Hall current. If the extrinsic
contribution to the Hall conductance is comparable with the intrinsic one,
at least a jump of the anomalous Hall conductance should be observed even if
the Hall current could not change its direction. This effect can be
identified as the intrinsic mechanism for anomalous Hall effect.

\subsection{2D Luttinger model}

We now turn our attention to the Luttinger Hamiltonian for spin $S=3/2$
holes in the valence band of centrosymmetric cubic semiconductors\cite%
{Luttinger56pr}%
\begin{equation}
H=\frac{1}{2m}\left( \gamma _{1}+\frac{5}{2}\gamma _{2}\right) p^{2}-\frac{%
\gamma _{2}}{m}\left( \mathbf{p}\cdot \mathbf{S}\right) ^{2},
\label{Luttinger}
\end{equation}%
where $\gamma _{1}$, $\gamma _{2}$ are material-dependent parameters and $m$
is the electron mass. In terms of $SO\left( 5\right) $ Clifford algebra, the
Hamiltonian (\ref{Luttinger}) can be cast into\cite{Murakami04prb}%
\begin{equation}
H=\epsilon \left( \mathbf{p}\right) +d_{a}\Gamma ^{a},
\end{equation}%
where $\epsilon \left( \mathbf{p}\right) =\gamma _{1}p^{2}/2m$, $\Gamma ^{a}$
$\left( a=1,2,\cdots ,5\right) $ the five Dirac $\Gamma $ matrices and $%
d_{a} $ the five $d$-wave combinations of $p$ are given in Ref.\cite%
{Murakami04prb}. In the two-dimensional case, for the first heavy- and
light-hole bands, the confinement in a well of thickness $a$ is approximated
by the relation $\left\langle p_{z}\right\rangle =0$, $\left\langle
p_{z}^{2}\right\rangle \simeq \left( \pi \hbar /a\right) ^{2}$.\cite%
{Bernevig05prl} In this case, $d_{1}=d_{2}=0$ and $\Gamma ^{a}\left(
a=3,4,5\right) $ forms reducible representation of an $SO\left( 3\right) $
Clifford sub-algebra.\cite{Qi05xxx}. In the following, one introduces a new
representation under which the expressions of $\mathbf{S}$ matrices and $%
\Gamma ^{a}$ ($a=3,4,5$) are given in Appendix. Under this new
representation, the two-dimension Luttinger model is block diagonal

\begin{equation}
H=\left(
\begin{array}{cc}
H_{+} & 0 \\
0 & H_{-}%
\end{array}%
\right) ,  \label{block}
\end{equation}%
where
\begin{equation}
H_{\mu }=\epsilon \left( p\right) +\mu \sum_{\alpha =x,y,z}d_{a}\sigma
_{\alpha },
\end{equation}%
with $\mu =\pm 1$, and
\begin{equation}
\epsilon \left( p\right) =\frac{\gamma _{1}}{2m}\left(
p_{x}^{2}+p_{y}^{2}\right) ,
\end{equation}%
\begin{equation}
d_{x}=-\frac{\sqrt{3}\gamma _{2}}{m}p_{x}p_{y},
\end{equation}%
\begin{equation}
d_{y}=-\frac{\sqrt{3}\gamma _{2}}{2m}\left( p_{x}^{2}-p_{y}^{2}\right) ,
\end{equation}%
\begin{equation}
d_{z}=-\frac{\gamma _{2}}{m}\left( \left\langle p_{z}^{2}\right\rangle -%
\frac{1}{2}p^{2}\right) .
\end{equation}%
(see Refs. \cite{Bernevig05prl,Kane05prl,Onada05prl}) These two effective
Hamiltonians are connected by the time reversal operator $\Theta =-i\sigma
_{y}K$ ($K$ is the complex conjugate operator that forms the complex
conjugate of any coefficient that multiplies a ket): $H_{+}=\Theta
H_{-}\Theta ^{-1}.$ Thus the energy eigenstates of $H$ are at least doubly
degenerated. $H_{\mu }$ has the same form as that for the ferromagnetic
metal with exchange coupling since $d_{z}(p)$ in $H_{\mu }$ contains a
non-zero term, $-\frac{\gamma _{2}}{m}\left\langle p_{z}^{2}\right\rangle $,
but the Pauli operators $\sigma _{\alpha }$ here do not represent a real
spin. To calculate the linear response of the spin current $J_{i}^{z}=\frac{1%
}{2}\left\{ \partial H/\partial p_{i},S_{z}\right\} $ to an electric current
$v_{j}=\partial H/\partial p_{j}$, the Kubo formula gives
\begin{equation}
\sigma _{xy}^{z}=\sum_{\mu =\pm 1}\left[ -2\sigma _{xy}^{z(\mu )}+\mu \frac{%
\hbar }{2e}\sigma _{xy}^{\left( \mu \right) }\right] .
\end{equation}%
where $\sigma _{xy}^{\left( \mu \right) }$ and $\sigma _{xy}^{z(\mu )}$ are
the anomalous and spin Hall conductance of $H_{\mu },$ respectively. The
anomalous Hall conductance $\sigma _{ij}=\sum_{\mu =\pm 1}\sigma
_{xy}^{\left( \mu \right) }.$ These relations are valid even if the
non-magnetic disorder and interaction are taken into account. It was known
that the spin Hall conductance is invariant $\sigma _{xy}^{z(+1)}=\sigma
_{xy}^{z(-1)}$, but the anomalous Hall conductance changes its sign $\sigma
_{xy}^{\left( +1\right) }=-\sigma _{xy}^{\left( -1\right) }$ under time
reversal. \textit{Therefore the calculation of spin Hall conductance in the
2D Luttinger model is reduced to those of spin Hall conductance and
anomalous Hall conductance in a ferromagnetic metal or band insulator of }$%
H_{\mu }.$ For a numerical calculation the material-specific parameters of
band structure for GaAs are adopted as $\gamma _{1}=6.92$ and $\gamma
_{2}=2.1$ the thickness of quantum well $a=8.3$ nm corresponds to the gap
between the light- and heavy-hole bands at the $\Gamma $-point $\Delta E=40$
meV. Using the formula of Eq. (\ref{ahc}) and (\ref{spinhall}) $\sigma
_{xy}^{\left( +1\right) }$, $\sigma _{xy}^{z(+1)}$, and $\sigma _{xy}^{z}$
are calculated numerically in terms of chemical potential $\mu $ as shown in
Fig. 2. It is noted that, for an infinite confinement, $\left\langle
p_{z}^{2}\right\rangle \rightarrow +\infty ,$ the eigenstates of $H_{\mu }$
becomes fully saturated.\cite{note2} As a result the spin-orbit coupling is
suppressed completely, and both spin and anomalous Hall effect vanish, which
is in agreement with Bernevig and Zhang.\cite{Bernevig05prl}

\begin{figure}[tbp]
\includegraphics[width=8cm]{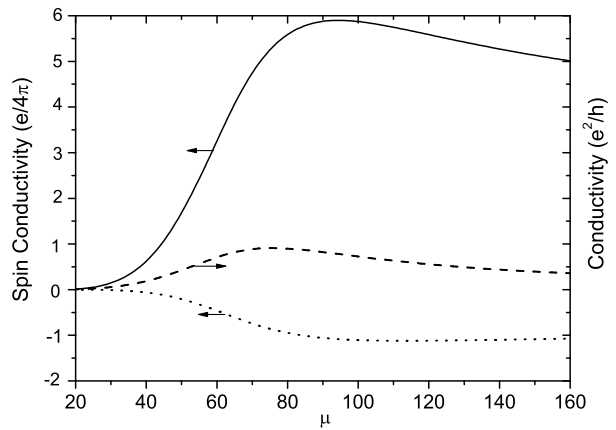}
\caption{Variations of $\protect\sigma _{xy}^{\left( +1\right) }$, $\protect%
\sigma _{xy}^{z(+1)} $, and $\protect\sigma _{xy}^{z}$ with Fermi level $%
\protect\mu $ (meV). The parameters are $a=8.3$ nm, $\protect\gamma _{1}=6.92
$ and $\protect\gamma _{2}=2.1$. The solid line corresponds to $\protect%
\sigma _{xy}^{z}$, the dotted line to $\protect\sigma _{xy}^{z\left(
+1\right) }$, and the dashed line to $\protect\sigma _{xy}^{\left( +1\right)
}$.}
\end{figure}

\section{Discussion}

Here we may present a heuristic picture for electronic transverse transports
in these systems and the effect of disorder from the point of view of spin
force. The effect of disorder on anomalous Hall conductance has been
investigated in Refs.\cite{Sinitsyn05prb,Dugaev05prb}. For simplicity we
focus on the system with the Wurtzite structure, i.e., $f=1$ in Eq.(\ref{f})
or $\beta =0$ in Eq.(\ref{RD}). The spin-orbit coupling induces the spin
force, which contains two parts: the transverse force on moving electron
spin,
\begin{equation}
\mathbf{F}_{1}=\frac{4m^{\ast 2}\lambda ^{2}}{\hslash ^{2}}J_{s}^{z}\times
\hat{z},
\end{equation}
\ and the exchange coupling interacting also induces a spin force within the
plane, which is relevant to the spin polarization,
\begin{equation}
\mathbf{F}_{2}=-\frac{2m^{\ast }\lambda h_{0}}{\hbar }\left[ \sigma _{x}\hat{%
x}+\sigma _{y}\hat{y}\right] .
\end{equation}
If the disorder potential $V_{disorder}$ is taken into account, in a steady
state, the spin force may reach at balance,
\begin{equation}
\frac{1}{i\hbar }\left\langle \left[ \frac{e}{c}\mathcal{A},H+V_{disorder}%
\right] \right\rangle =\left\langle \mathbf{F}_{1}+\mathbf{F}%
_{2}\right\rangle =0.
\end{equation}%
This result is independent of the non-magnetic disorder and interaction
because the spin gauge field commutes with non-magnetic potential $%
V_{disorder}$. From the spin force balance we have a relation between spin
current and spin polarization,
\begin{eqnarray}
\left\langle J_{x}^{z}\right\rangle &=&+\frac{\hbar h_{0}}{2m^{\ast }\lambda
}\left\langle \sigma _{y}\right\rangle , \\
\left\langle J_{y}^{z}\right\rangle &=&-\frac{\hbar h_{0}}{2m^{\ast }\lambda
}\left\langle \sigma _{x}\right\rangle .
\end{eqnarray}%
It becomes obvious that the spin Hall current vanishes in the case of $%
h_{0}=0$, which is consistent with previous complicated calculations.\cite%
{Inoue04prb,Mischenko04prl,Normura05prb,Adagideli05} Of course the purely
intrinsic responses of spin current and spin polarization without impurities
do not satisfy this relation, and the extrinsic contributions have to be
included to reach the balance. Thus, this gives a clear picture for
anomalous Hall effect in ferromagnetic metal: When the external electric
field is applied along the $x$ axis, it will circulate an electric current $%
J_{c,x}$, and also a spin current $J_{x}^{z}$ since the charge carriers are
partially polarized. The spin-orbit coupling exerts a spin transverse force
on the spin current, $J_{x}^{z}$, and generate a drift velocity or the
anomalous Hall current $J_{c,y}.$ From the Rashba coupling the spin
polarization tends to be normal to the momentum or electric current. The
electric current $J_{c,x}$ along $E$ induces a non-zero $\left\langle \sigma
_{y}\right\rangle $ and the anomalous Hall current $J_{c,y}$ induces
non-zero $\left\langle \sigma _{x}\right\rangle .$ These non-zero spin
polarization maintains the balance of spin transverse force, and further a
non-zero spin current in a steady state. Thus the anomalous electronic
transverse transport is robust against the disorder in the ferromagnetic
metals and semiconductors. This picture can also be applied to understand
the spin Hall effect in the quasi-2D Luttinger system.

\section*{Acknowledgments}

This work was supported by the Research Grant Council of Hong Kong under
Grant No.: HKU 7039/05P (SQS), and Hubei Natural Science Foundation of China
under Grant No. 2005ABA307 (BZ).

\section*{Appendix}

In order to write the two-dimensional Hamiltonian in the form of Eq. (\ref%
{block}), one introduces a new representation under which $\mathbf{S}$
matrices are expressed as$\ $
\begin{equation}
S_{z}=\left(
\begin{array}{cccc}
-\frac{3}{2} & 0 & 0 & 0 \\
0 & \frac{1}{2} & 0 & 0 \\
0 & 0 & -\frac{1}{2} & 0 \\
0 & 0 & 0 & \frac{3}{2}%
\end{array}%
\right) ,  \tag{A1}
\end{equation}%
\begin{equation}
S_{x}=\left(
\begin{array}{cccc}
0 & 0 & -i\frac{\sqrt{3}}{2} & 0 \\
0 & 0 & 1 & i\frac{\sqrt{3}}{2} \\
i\frac{\sqrt{3}}{2} & 1 & 0 & 0 \\
0 & -i\frac{\sqrt{3}}{2} & 0 & 0%
\end{array}%
\right) ,  \tag{A2}
\end{equation}%
\begin{equation}
S_{y}=\left(
\begin{array}{cccc}
0 & 0 & \frac{\sqrt{3}}{2} & 0 \\
0 & 0 & -i & -\frac{\sqrt{3}}{2} \\
\frac{\sqrt{3}}{2} & i & 0 & 0 \\
0 & -\frac{\sqrt{3}}{2} & 0 & 0%
\end{array}%
\right) .  \tag{A3}
\end{equation}%
Now the Dirac matrices $\Gamma ^{a}$ ($a=3,4,5$) become%
\begin{equation}
\Gamma _{3}=\frac{1}{\sqrt{3}}\left( S_{x}S_{y}+S_{y}S_{x}\right) =\sigma
_{z}\otimes \sigma _{x},  \tag{A4}
\end{equation}%
\begin{equation}
\Gamma _{4}=\frac{1}{\sqrt{3}}\left( S_{x}^{2}-S_{y}^{2}\right) =\sigma
_{z}\otimes \sigma _{y},  \tag{A5}
\end{equation}%
\begin{equation}
\Gamma _{5}=S_{z}^{2}-\frac{5}{4}=\sigma _{z}\otimes \sigma _{z},  \tag{A6}
\end{equation}%
and furthermore
\begin{equation}
\Gamma _{12}=\frac{1}{2i}\left[ \Gamma _{1},\Gamma _{2}\right] =\sigma
_{z}\otimes \mathbf{I,}  \tag{A7}
\end{equation}%
\begin{equation}
\Gamma _{34}=\frac{1}{2i}\left[ \Gamma _{3},\Gamma _{4}\right] =\mathbf{I}%
\otimes \mathbf{\mathbf{\sigma }_{z},}  \tag{A8}
\end{equation}%
\begin{equation}
S_{z} =-\Gamma _{34}-\frac{1}{2}\Gamma _{12}.  \tag{A9}
\end{equation}


\begin{thebibliography}{99}
\bibitem{AHE} C.M. Hurd, \textit{The Hall Effect in Metals and Alloys}
(Plenum, New York, 1972); \textit{The Hall Effect and Its Applications},
edited by C.L. Chien and C.R. Westgate (Plenum, New York, 1980).

\bibitem{Karplus54pr} R. Karplus and J.M. Luttinger, Phys. Rev. \textbf{95},
1154 (1954).

\bibitem{Luttinger58pr} J.M. Luttinger, Phys. Rev. \textbf{112}, 739 (1958).

\bibitem{Smit58p} J. Smit, Physica (Amsterdam) \textbf{24}, 39 (1958).

\bibitem{Berger70prb} L. Berger, Phys. Rev. B \textbf{2}, 4559 (1970).

\bibitem{Jungwirth02prl} T. Jungwirth, Q. Niu, and A.H. MacDonald, Phys.
Rev. Lett. \textbf{88}, 207208 (2002).

\bibitem{Onada02jpsj} M. Onada and N. Nagaosa, J. Phys. Soc. Jpn. \textbf{71}%
, 19 (2002).

\bibitem{Murakami03sci} S. Murakami, N. Nagaosa, and S.C. Zhang, Science
\textbf{301}, 1348 (2003).

\bibitem{Sinova04prl} J. Sinova, D. Culcer, Q. Niu, N.A. Sinitsyn,
T.Jungwirth, and A.H. MacDonald, Phys. Rev. Lett. \textbf{92}, 126603 (2004).

\bibitem{Shen04prl} S.Q. Shen, M. Ma, X.C. Xie, and F.C. Zhang, Phys. Rev.
Lett. \textbf{92}, 256603 (2004); S.Q. Shen, Y.J. Bao, M. Ma, X.C. Xie, and
F.C. Zhang, Phys. Rev. B \textbf{71}, 155316 (2005).

\bibitem{Fang04Science} Z. Fang, N. Nagaosa, K.S. Takahashi, A. Asamitsu, R.
Mathieu, T. Ogasawara, H. Yamada, M. Kamasaki, Y. Tokura, and K. Terakura,
Science \textbf{302}, 92 (2003).

\bibitem{Lee04sci} W.-L. Lee, S. Watauchi, V.L. Miller, R.J. Cava, and N.P.
Ong, Science \textbf{303}, 1647 (2004).

\bibitem{Wunderlich05prl} J. Wunderlich, B. Kaestner, J. Sinova, and T.
Jungwirth, Phys. Rev. Lett. \textbf{94}, 047204 (2005).

\bibitem{Shen05prl} S.Q. Shen, Phys. Rev. Lett. \textbf{95}, 187203 (2005);
J. Li, L. Hu, and S.Q. Shen, Phys. Rev. B \textbf{71}, 241305(R) (2005).

\bibitem{Mahan} G.D. Mahan, Many-Particle Physics, (Plenum, New York, 1990).

\bibitem{singularity} If $d=0$ at some points, usually we cannot take the
limit of $\delta^{+} \rightarrow 0^{+}$ before the integral over $p$.

\bibitem{Qi05xxx} X.L. Qi, Y.S. Wu and S.C. Zhang, cond-mat/0505308.

\bibitem{Bernevig05prb} B.A. Bernevig, Phys. Rev. B, \textbf{71}, 073201(R)
(2005).

\bibitem{Sundaram99prb} G. Sundaram and Q. Niu, Phys. Rev. B \textbf{59},
14915 (1999).

\bibitem{Culcer03prb} D. Culcer, A. MacDonald, and Q. Niu, Phys. Rev. B
\textbf{68}, 045327 (2003).

\bibitem{Shen04prb} S.Q. Shen, Phys. Rev. B, \textbf{70}, 081311(R) (2004).

\bibitem{Luttinger56pr} J.M. Luttinger, Phys. Rev. \textbf{102}, 1030 (1956).

\bibitem{Murakami04prb} S. Murakami, N. Nagaosa, and S.C. Zhang, Phys. Rev.
B \textbf{69}, 235206 (2004).

\bibitem{Bernevig05prl} B.A. Bernevig and S.C. Zhang, Phys. Rev. Lett.
\textbf{95}, 016801 (2005).

\bibitem{Kane05prl} C.L. Kane and E.J. Mele, Phys. Rev. Lett. \textbf{95},
146802 (2005).

\bibitem{Onada05prl} M. Onoda and N. Nagaosa, Phys. Rev. Lett. \textbf{95},
106601 (2005).

\bibitem{note2} On a lattice model, numerical value is slightly revised.

\bibitem{Sinitsyn05prb} N.A. Sinitsyn, Q. Niu, J. Sinova, and K. Nomura,
Phys. Rev. B \textbf{72}, 045346 (2005).

\bibitem{Dugaev05prb} V.K. Dugaev, P. Bruno, M. Taillefumier, B. Canals, and
C. Lacroix, Phys. Rev. B \textbf{71}, 224423 (2005).

\bibitem{Inoue04prb} J.I. Inoue, G.E. W. Bauer and L.W. Molenkamp, Phys.
Rev. B \textbf{70}, 041303(R) (2004).

\bibitem{Mischenko04prl} E.G. Mishchenko, A.V. Shytov, and B.I. Halperin,
Phys. Rev. Lett. \textbf{93}, 226602 (2004).

\bibitem{Normura05prb} K. Nomura, J. Sinova, T. Jungwirth, Q. Niu, and A.H.
MacDonald, Phys. Rev. B \textbf{71}, 041304(R) (2005).

\bibitem{Adagideli05} \.{I}. Adagideli and G.E.W. Bauer, Phys. Rev. Lett.
\textbf{95}, 256602 (2005).
\end{thebibliography}
\end{document}